\documentclass[sigconf,natbib=true,anonymous=false]{acmart}
\usepackage{booktabs} 
\usepackage{multirow}
\usepackage{array}
\usepackage{graphicx}

\usepackage[normalem]{ulem}
\useunder{\uline}{\ul}{}
\usepackage{subcaption}
\usepackage{cleveref}
\usepackage{rotating}
\usepackage{siunitx}
\usepackage{makecell}
\usepackage[linesnumbered,ruled] {algorithm2e} 
\usepackage{tablefootnote}

\usepackage{framed}
\usepackage{color}
\definecolor{shadecolor}{rgb}{0.92,0.92,0.92}

\usepackage{listings}
\usepackage{xcolor}

\colorlet{punct}{red!60!black}
\definecolor{background}{HTML}{EEEEEE}
\definecolor{delim}{RGB}{20,105,176}
\colorlet{numb}{magenta!60!black}

\lstdefinelanguage{json}{
    basicstyle=\normalfont\ttfamily,
    numbers=left,
    numberstyle=\scriptsize,
    stepnumber=1,
    numbersep=8pt,
    showstringspaces=false,
    breaklines=true,
    frame=lines,
    backgroundcolor=\color{background},
    literate=
     *{0}{{{\color{numb}0}}}{1}
      {1}{{{\color{numb}1}}}{1}
      {2}{{{\color{numb}2}}}{1}
      {3}{{{\color{numb}3}}}{1}
      {4}{{{\color{numb}4}}}{1}
      {5}{{{\color{numb}5}}}{1}
      {6}{{{\color{numb}6}}}{1}
      {7}{{{\color{numb}7}}}{1}
      {8}{{{\color{numb}8}}}{1}
      {9}{{{\color{numb}9}}}{1}
      {:}{{{\color{punct}{:}}}}{1}
      {,}{{{\color{punct}{,}}}}{1}
      {\{}{{{\color{delim}{\{}}}}{1}
      {\}}{{{\color{delim}{\}}}}}{1}
      {[}{{{\color{delim}{[}}}}{1}
      {]}{{{\color{delim}{]}}}}{1},
}

\usepackage{amssymb}
\usepackage{pifont}

\newcommand{\argmax}{\arg\!\max}

\newcolumntype{S}{>{\centering\arraybackslash}m{2cm}}
\newcolumntype{M}{>{\centering\arraybackslash}m{3cm}}
\newcolumntype{L}{>{\centering\arraybackslash}m{4cm}}

\begin{document}

\title[PRE: A Peer Review Based Large Language Model Evaluator
]{PRE: A Peer Review Based Large Language Model Evaluator
}

\author{Zhumin Chu}
\affiliation{%
  \institution{Quan Cheng Laboratory, \\
DCST, Tsinghua University\\
Zhongguancun Laboratory}
  \city{Beijing}
  \country{China}
  \postcode{100084}
}
\email{chuzm19@mails.tsinghua.edu.cn}

\author{Qingyao Ai\footnotemark}
\affiliation{%
  \institution{Quan Cheng Laboratory, \\
DCST, Tsinghua University\\
Zhongguancun Laboratory}
  \city{Beijing}
  \country{China}
  \postcode{100084}
}
\email{aiqy@tsinghua.edu.cn}

\author{Yiteng Tu}
\affiliation{%
  \institution{Quan Cheng Laboratory, \\
DCST, Tsinghua University\\
Zhongguancun Laboratory}
  \city{Beijing}
  \country{China}
  \postcode{100084}
}
\email{yitengtu16@gmail.com}

\author{Haitao Li}
\affiliation{%
  \institution{Quan Cheng Laboratory, \\
DCST, Tsinghua University\\
Zhongguancun Laboratory}
  \city{Beijing}
  \country{China}
  \postcode{100084}
}
\email{liht22@mails.tsinghua.edu.cn}

\author{Yiqun Liu}
\affiliation{%
  \institution{Quan Cheng Laboratory, \\
DCST, Tsinghua University\\
Zhongguancun Laboratory}
  \city{Beijing}
  \country{China}
  \postcode{100084}
}
\email{yiqunliu@tsinghua.edu.cn}

\begin{abstract}
The impressive performance of large language models (LLMs) has attracted considerable attention from the academic and industrial communities. Besides how to construct and train LLMs, how to effectively evaluate and compare the capacity of LLMs has also been well recognized as an important yet difficult problem. Existing paradigms rely on either human annotators or model-based evaluators to evaluate the performance of LLMs on different tasks. However, these paradigms often suffer from high cost, low generalizability, and inherited biases in practice, which make them incapable of supporting the sustainable development of LLMs in long term. In order to address these issues, inspired by the peer review systems widely used in academic publication process, we propose a novel framework that can automatically evaluate LLMs through a peer-review process. Specifically, for the evaluation of a specific task, we first construct a small qualification exam to select ``reviewers'' from a couple of powerful LLMs. Then, to actually evaluate the ``submissions" written by different candidate LLMs, i.e., the evaluatees, we use the reviewer LLMs to rate or compare the submissions. The final ranking of evaluatee LLMs is generated based on the results provided by all reviewers. We conducted extensive experiments on both text summarization and non-factoid question answering tasks with eleven LLMs including GPT-4. The results demonstrate the existence of biasness when evaluating using a single LLM. Also, our PRE model outperforms all the baselines, illustrating the effectiveness of the peer review mechanism.
\end{abstract}

%

%
%


\renewcommand\footnotetextcopyrightpermission[1]{} 
\settopmatter{printacmref=false}
\maketitle


\section{Introduction}
\label{sec:introduction}
 
The continuous development of large-scale language models (LLMs) such as GPT-3~\cite{brown2020language}, PALM~\cite{chowdhery2022palm}, and Llama~\cite{llama} has sparked people's passion on Artifical General Intelligence in both academia and industry. A new generation of LLMs, led by GPT-4~\cite{gpt4} and Claude~\cite{claude, claude2}, can achieve competitive performance on a wide range of natural language processing tasks, even in zero-shot scenarios. Ever since the release of ChatGPT~\cite{chatgpt}, a large number of LLMs have been developed, many of which can produce high-quality responses that achieve or even surpass human-level performance in many cases~\cite{mao2023gpteval, ali2022performance}. 

With rapid development of LLMs, how to evaluate the performance of LLMs both effectively and efficiently has become a crucial bottleneck that restricts LLMs' progress. A reliable and reusable LLM evaluation method not only helps us better select the best LLMs for each task, but also provides important guidelines for LLM optimization. 

To the best of our knowledge, there are two types of evaluation paradigms for LLM: human evaluation and model-based evaluation. 
The former hires human annotators to judge the quality of responses generated by LLMs directly or create gold references to evaluate the outputs of LLMs. The later trains separate evaluators for each task or uses a powerful LLM (e.g., GPT-4) to evaluate the performance of other LLMs.
Unfortunately, due to their intrinsic characteristics, these methods often suffer from one or more of the following three problems:

(1) \textbf{High cost}: Human annotations have been considered the most effective and reliable data to evaluate the quality of LLM outputs~\cite{zheng2023judging, frieder2023mathematical, jang2022becel}. However, in commonly-used generation tasks, such as text summarization and question answering, different LLMs would output diverse responses, leading the cost of evaluation be approximately proportional to the number of evaluated LLMs. Reference-based methods~\cite{rouge, bleu, bertscore} try to avoid this problem by requiring the annotators to provide gold references for each tasks instead of judging the quality of each LLM's outputs directly, but this could significantly increase the load and difficulty of the annotation jobs. Also, since LLMs are extremely powerful in terms of memorization, any public reference-based datasets can easily be incorporated and optimized by LLMs in the training process and thus become useless for evaluation after a short period of time. All these make the cost of annotation-based LLM evaluation methods prohibitive in long term.
    
(2) \textbf{Low generalizibility}: Existing evaluation methods, such as reference-based or model-based evaluators, often requires task-specific dataset construction and evaluator pre-training~\cite{superclue, safetyprompts, kryscinski2019evaluating, hendrycks2020measuring, gu2023xiezhi}. For example, Xu et al.~\cite{superclue} designed multiple-choice questions based on human references to evaluate LLMs. Similarly, studies like~\cite{safetyprompts} fine-tune pretrained language models on each specific task with large-scale supervised data to create model-based evaluators. However, the evaluators created in these methods cannot be generalized to tasks beyond the target task of the references or the training data. Considering the large number and variety of LLM applications, the low generalizability of these evaluation methods make them not preferable for LLM evaluation.

(3) \textbf{Inherent bias}: Due to their intrinsic model structure or algorithm design, many evaluation methods are inherently biased in the evaluation process. For example, reference-based word similarity metrics (such as ROUGE~\cite{rouge}, BLEU~\cite{bleu}, and BERTScore~\cite{bertscore}), which are commonly used to evaluate the outputs of LLMs in generation tasks, steer LLM outputs to be as similar as possible to the reference text, discriminating against LLMs that create qualified but different responses. Recently, many studies have adopted the state-of-the-art LLM, GPT-4, as their evaluation tools~\cite{liu2023gpteval, kocmi2023large}. Although several works have demonstrated that GPT-4 has decent evaluation capabilities~\cite{liu2023gpteval, kocmi2023large}, we found that GPT-4, so as other LLMs, often prefers responses of LLMs from its own series (i.e., the GPT models) over other LLMs despite of the actual quality of the responses. In other words, if we use GPT-4 as the evaluator, its inherent bias may make it difficult, if possible, to develop an LLM outside the GPT family that outperforms GPT-4.

To address the aforementioned issues, we propose a novel framework, Peer Review Evaluator (PRE)~\footnote{\url{https://anonymous.4open.science/r/PRE-D66A}}, to evaluate the performance of LLMs automatically. Inspired by the peer review mechanism in academic community, we propose to use LLMs as reviewers to evaluate the performance of LLMs directly. 
Specifically, we first develop a qualification exam to filter out LLMs that fail to provide reliable evaluation results. Then, qualified reviewer LLMs are required to assess the outputs of the evaluatee LLMs, and the final evaluation results are aggregated from all reviewer LLMs' ratings or preferences. To verified the effectiveness of our framework, we conducted extensive experiments on two representative text generation tasks, i.e., document summarization and non-factoid question answering. The experimental results show that the results of PRE model have the highest consistency with human preferences (ground truth) compared to all the baseline models including GPT-4. Comparing to previous evaluation methods, PRE can easily be generalized to different tasks and is highly cost efficient. Also, experiment results show that PRE provides much more robust evaluation results than methods that rely on specific model structures or LLMs.

 In summary, our contributions can be summarized as follows:
 \begin{itemize}
 	\item We propose a novel and automatic LLM evaluation framework PRE that incorporates peer review mechanisms.
 	\item Through the use of qualification exams and result fusions, PRE can achieve effective LLM evaluation while being robust to potential model bias, which has been widely observed in existing automatic evaluation methods.
 	\item We conducted extensive experiments with both the document summarization and non-factoid QA task to demonstrate the potential of PRE.
 \end{itemize}

\vspace{-3mm}
\section{Related Work}
\label{sec:related work}

\subsection{Large Language Models}
Large Language Models (LLMs) typically refer to language models that contain more than a hundred billion parameters and have been pre-trained on large amounts of textual data. The large-scale parameters and large amounts of training data of LLMs bring impressive capabilities, such as few-shot and zero-shot learning, where they can generate high-quality and reasonable text output with limited prompts. 
 In addition, LLMs offer emergent abilities~\cite{wei2022emergent} that are not observed in smaller models, which are reflected in their stunning generalization performance on unseen tasks.

According to open source or not, LLMs can be divided into two categories: closed source LLMs and open source LLMs. Closed source LLMs include ChatGPT~\cite{chatgpt}, GPT-4~\cite{gpt4}, Claude~\cite{claude}, Claude 2~\cite{claude2} and Gemini~\cite{team2023gemini} which only offer API services instead of publicly available models. They tend to have enormous parameter sizes, and therefore reach top performance on all types of tasks. 

For open source models, 
 LLaMa is a foundation language model trained on publicly available data with parameter counts from 7B to 65B, which has shown decent performance on various benchmarks. Due to its excellent performance and the convenience of open source, many researchers have built on it to conduct continual pre-training or instruction tuning so as to enhance its capabilities further. Such customized models include Alpaca~\cite{alpaca}, Koala~\cite{koala}, and Vicuna~\cite{vicuna}. ChatGLM and ChatGLM2~\cite{glm}, in addition to instruction tuning, are committed to utilizing quantitative methods to reduce model memory footprint and improve inference efficiency. FastChat-T5~\cite{fastt5} is based on the encoder-decoder transformer architecture, and is fine-tuned on the basis of Flan-T5~\cite{flant5} with user-shared conversation data. Baichuan~\cite{baichuan} is a Chinese-English bilingual language model trained on about 1.2 trillion tokens. Unlike the previous models, RWKV~\cite{rwkv} adopts recurrent neural networks (RNNs)~\cite{rumelhart1986learning} as its underlying architecture. It can significantly reduce computational resources and improve computational efficiency.
 
\vspace{-3mm}
\subsection{Evaluation of Large Language Models}
With the rapid development of LLMs, how to effectively evaluate the quality of the LLM generative texts has also become an urgent research question. We can simply categorize the existing evaluation approaches into the following categories:

\textbf{Human Annotations}: Human annotation has usually been regarded the most effective and reliable means for evaluating the outputs of LLMs. Recently, LMSYS has built a benchmark platform, Chatbot Arena~\cite{zheng2023judging, frieder2023mathematical, jang2022becel}, which allows different LLMs to engage in a fair, anonymous and random battle through crowdsourcing manner. Then, they adopted the ELO rating system to aggregate for the final leaderboard. However, as the number of evaluatee LLMs and evaluation tasks sharply increase, human annotation becomes increasingly unsustainable. Thus, effective semi-automatic or automatic evaluation methods become an urgent need.

\textbf{Reference-based Word Similarity Metrics}: Before the emergence of LLMs, there are a number of similarity metrics that could assess the quality of the generative text based on the reference text. 
 BLEU~\cite{bleu} is a simple evaluation metric that focuses on the n-gram matches between generative text and reference text. ROUGE~\cite{rouge} is a set of recall-oriented metrics that measures the number of matching units like n-gram, word sequences, and word pairs. BERTScore~\cite{bertscore} is an embedding-based metric that maps texts to embedding vectors and evaluates the quality of the generative text via cosine similarity.

\textbf{Evaluation with Multi-Choice Questions}: Multiple-choice questions, as a category of questions with fixed output formats, whose easy-evaluation properties lead a great deal of work~\cite{superclue,gaokao} to construct benchmarks with such formatting. Such evaluation results are often more intuitive and aligned with human values. 
 SuperCLUE~\cite{superclue} is a Chinese comprehensive benchmark that assesses LLMs' general competencies through multiple-choice questions. GAOKAO~\cite{gaokao} selects questions from the Chinese college entrance exams to assess the language comprehension and logical reasoning abilities of LLMs.

\textbf{Evaluation using LLMs}: Due to the stunning performance of LLMs, many studies attempted to employ one or multiple LLMs as evaluators for the evaluation of LLMs' outputs. GPTScore\cite{gptscore} evaluates the quality of generative texts using the generation probability of LLMs. 
It argues that higher quality text is more likely to be generated with the given instructions and context. 
PandaLM~\cite{pandalm} trains LLaMa~\cite{llama} to evaluate the results generated by LLMs through instruction tuning. Its training data is generated by ChatGPT via self-instruct~\cite{selfinstruct}. 
 Safety-Prompts~\cite{safetyprompts} is a Chinese LLMs safety assessment benchmark that utilizes LLMs to detect the potential unsafe situations in the input texts. 
PRD~\cite{li2023prd} and CHATEVAL~\cite{chan2023chateval}, the two recent works, attempt to integrate multiple LLMs into the evaluation system to provide an aligned evaluation result by ranking, discussing, and debating among LLMs. 
 Their experimental results found that a synergy of multiple LLMs could produce a higher consistency of evaluation results with human judgments.
 Different from previous studies, in this work, we propose an innovative evaluation framework for large language models based on the peer review system. This framework can automatically and unbiasedly integrate the evaluation results of multiple LLMs, providing more reliable assessments.

\begin{figure*}[thbp]
   \captionsetup[subfigure]{justification=centering}
   \centering
 \includegraphics[width=0.8\textwidth]{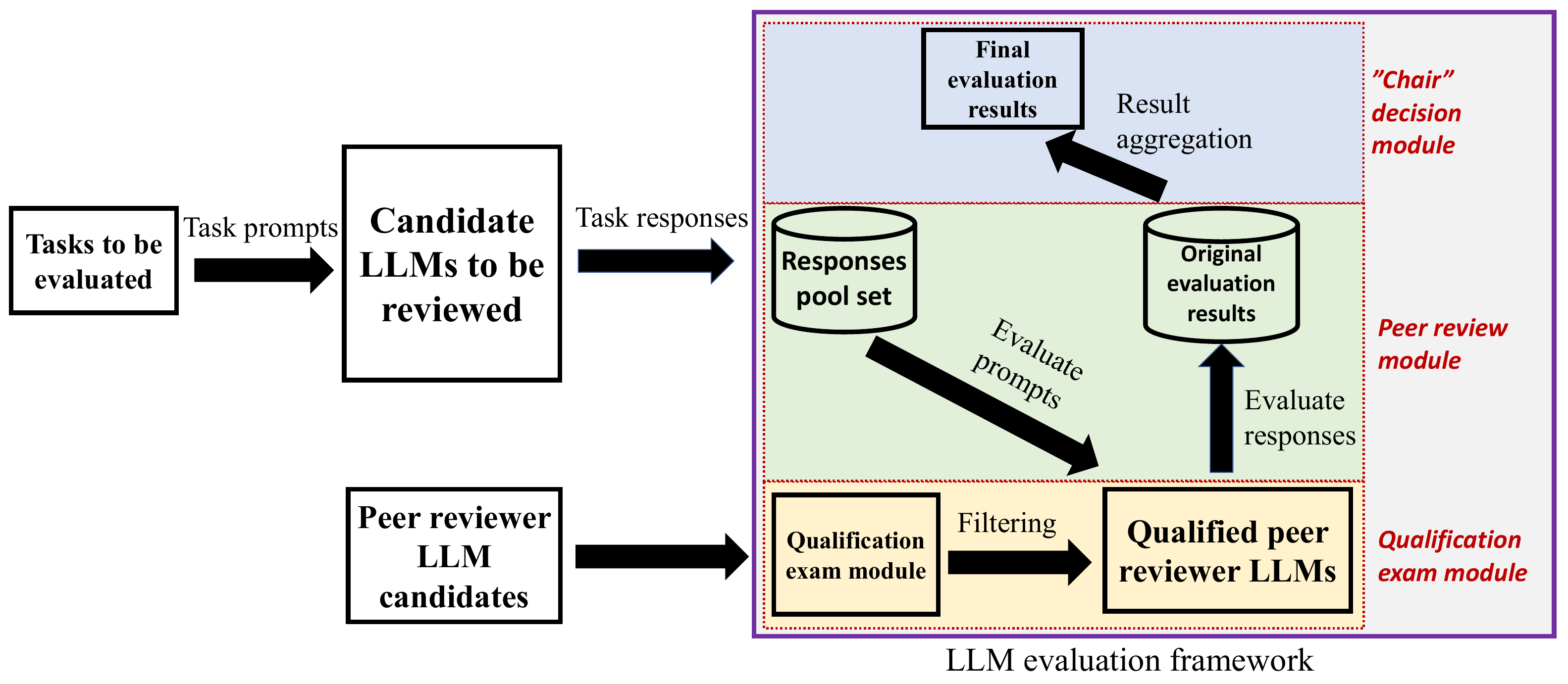}
  \vspace{-3mm}
   \caption{The architecture of our evaluation framework for large language models}
   \label{fig:The architecture of our evaluation framework for large language models}
 \end{figure*}
 
\section{LLM Peer Review Evaluation}
\label{sec:Framework Design}

In this section, we provide a detailed introduction to the design motivation and specifics of our proposed LLMs evaluation framework.

\vspace{-3mm}
\subsection{Motivation}
As discussed in Sec~\ref{sec:introduction}, a good LLM evaluation system should be affordable, generalizable and unbiased. Existing evaluation methods powered by a strong LLM (e.g., GPT-4) have been shown to be both effective and cost-efficient~\cite {mao2023gpteval, ali2022performance}, but they also suffer from intrinsic limitations like inherent bias as discussed in Section~\ref{sec:introduction}. To this end, we propose to employ the peer review mechanism to integrate the evaluation results of multiple LLMs.
 
Peer review mechanisms are widely used in the academic field for paper reviewing. Journal editors or conference chairs invite experienced researchers in such research fields to act as reviewers, providing feedback and ratings on submitted papers. Editors or chairs take into account the reviewers' comments to make final decisions. Inspired by this mechanism, we apply it to the scenario of LLMs evaluation. Specifically, we consider multiple LLMs as potential reviewers. The evaluation framework, acting as the chair, selects qualified LLM reviewers to rate the outputs of each LLM on the task, and ultimately aggregates the reviewers' rates to provide the final evaluation results.

\vspace{-3mm}
\subsection{Framework Architecture}
\label{subsec:framework architecture}

Figure~\ref{fig:The architecture of our evaluation framework for large language models} shows the architecture of our overall LLM evaluation framework: Peer Review Evaluator (PRE). The whole process can be divided into three modules: (1) Qualification exam module: conducting a qualification exam for all reviewer candidate LLMs to select qualified LLMs exceeding a certain level of evaluation capability; (2) Peer review module: collecting the outputs of evaluatee LLMs on the given assessment tasks, and then rating the outputs of all evaluatee LLMs by qualified reviewer LLMs; (3) ``Chair'' decision module: aggregating the ratings provided by all reviewer LLMs to obtain the final evaluation results. Below, we will provide detailed information regarding the design details of each module.

\vspace{-2mm}
\subsubsection{Qualification exam module}
\begin{figure}[t]

   \captionsetup[subfigure]{justification=centering}
   \centering
 \includegraphics[width=0.45\textwidth]{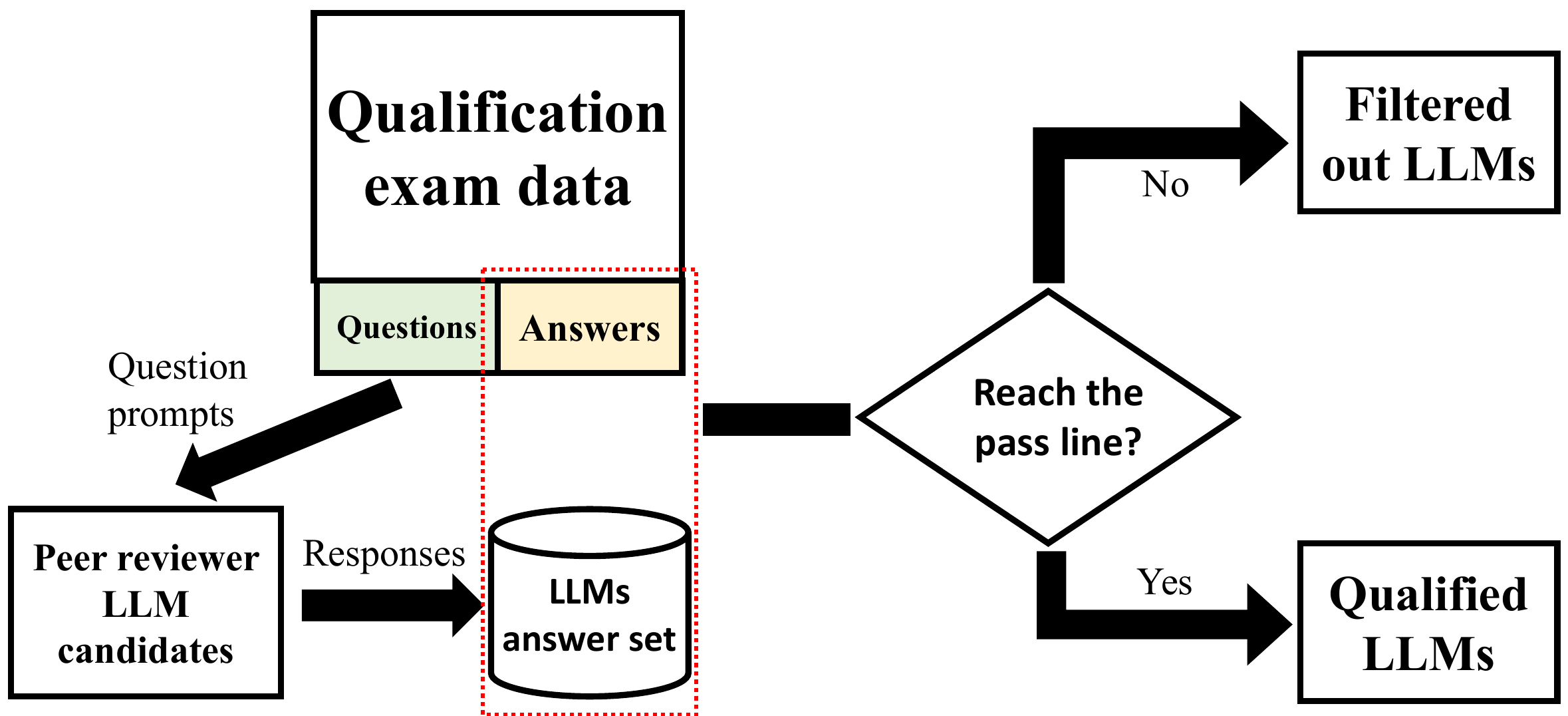}
  \vspace{-3mm}
   \caption{The process of the qualification exam module in our evaluation framework}
   \label{fig:The specific process of the qualification exam module in our evaluation framework}
  \vspace{-3mm}
 \end{figure}

Previous work has already demonstrated that LLMs have certain evaluation capabilities~\cite{mao2023gpteval, ali2022performance}. Based on this finding, in our framework, any LLMs are allowed to participate in the evaluation process as reviewer candidates. Through the qualification examination module, we select LLMs whose evaluation capability is strong enough from the reviewer candidates to participate as reviewers in the peer review stage. Figure~\ref{fig:The specific process of the qualification exam module in our evaluation framework} illustrates the specific process of the qualification exam. This process relies on a set of qualification examination data, which should include a set of test cases to assess the evaluation capability of LLMs. We require each reviewer candidate to complete these evaluation tasks, and then compare candidates' outputs with the standard answers to obtain the evaluation scores for each candidate's capability. Only when the evaluation score of a reviewer candidate LLM reaches the admission threshold, will we add it to the reviewer pool.

 There are two points worth further discussion regarding the qualification exam data: (1) Data acquisition: In order to closely approximate the application scenarios of reviewer LLMs, in our experimental setup, we used the outputs of a subset of LLMs in the evaluated task as the evaluation objects, constructing the qualification exam data. For simplicity, we use human annotations to create the qualification exam data (described in Sec~\ref{subsec:Manual Annotations}), but please note that other unsupervised or semi-supervised methods~\cite{yue2023automatic, kryscinski2019evaluating} could also be used to create the exam. (2) Data reusability: The purpose of the qualification exam data is to assess the evaluation abilities of reviewer candidate LLMs. With proper design, a single set of qualification exam data can reflect the general evaluation abilities of reviewer candidate LLMs, thus making the reviewer selection results generalizable to multiple tasks. Also, the exam data is designed to evaluate LLM's ability as a reviewer. After the exam data are published, even if an LLM manages to trick the exam and become a reviewer by using the data in training, it doesn't mean that the LLM could stand out as an evaluatee in the actual testing stage (i.e., the peer reviewing process). This makes the whole framework more robust and reusable.  
 
 \vspace{-2mm}
 \subsubsection{Peer review module}
 In this module, we first collect the responses of all evaluatee LLMs to the given tasks. Then, each qualified reviewer LLM is required to rate the outputs in the response pool set. Specifically, organizers need to design prompts in advance for rating, and feed them into the reviewer LLMs. Then, they extract corresponding rating information from the reviewers' outputs. It is worth noting that the rating method here is not limited to pointwise evaluation of each (task, response) pair, but can also be in pairwise or even listwise format.
 
 \vspace{-2mm}
 \subsubsection{``Chair'' decision module}
 \label{subsec:chair decision module}
 After collecting the comments (ratings) from all the LLM reviewers, the ``chair'' (evaluation system) needs to aggregate the ratings to generate the final evaluation results. Specifically, we adopt a weighted voting strategy for rating aggregation, as shown in Eq~\ref{eq:weighted vote strategy}. 
 
  \begin{equation}
 	R_x = \frac{1}{W}\sum_{l\in L} w_l r_x^{(l)}
 	\label{eq:weighted vote strategy}
 \end{equation}

 In Eq~\ref{eq:weighted vote strategy}, $L$ denotes the whole reviewer LLM set, and $r_x^{(l)}$ denotes the LLM $l$'s rating on sample $x$. The vote weight $w_l$ of each reviewer LLM is determined by its performance in the qualification exam, while  $W$ is the normalization term with $W = \sum_{l\in L} w_l$.

\section{Experimental Setup}
\label{sec:Experimental Setup}

\begin{table*}[]
\centering
\caption{The basic information of the large language models used in our experiments}
\label{tab:The basic information of the large language models used in our experiments}
\begin{tabular}{lccp{3cm}<{\centering}cp{1cm}<{\centering}p{1cm}<{\centering}p{1cm}<{\centering}}
\toprule[1.0pt]	
Model & Developer & Size (B) & ELO rate (rank) & Evaluatee & Evaluator   candidate & Annotation & Exam provider \\
\midrule[0.6pt]
GPT-4~\cite{gpt4}            & Openai        & /  & 1193 (1 / 28)      & \checkmark & \checkmark &   &   \\
Claude-1~\cite{claude}         & Anthropic     & /  & 1161 (2 / 28)      & \checkmark & \checkmark & \checkmark &   \\
GPT-3.5-turbo~\cite{chatgpt}    & Openai        & /  & 1118 (5 / 28)      & \checkmark & \checkmark & \checkmark & \checkmark \\
Llama-2-70b-chat~\cite{touvron2023llama} & Meta          & 70 & 1060 (7 / 28)      & \checkmark & \checkmark &   &   \\
Vicuna-7b~\cite{vicuna}         & LMSYS         & 7  & 1003 (14 / 28)     & \checkmark & \checkmark & \checkmark &   \\
ChatGLM2-6B~\cite{glm}      & Tsinghua      & 6  & 965 (18 / 28)      & \checkmark & \checkmark & \checkmark &   \\
RWKV-4-Raven-7B~\cite{rwkv}  & BlinkDL       & 7  & 14B: 939 (21 / 28) & \checkmark & \checkmark & \checkmark &   \\
Alpaca-7b~\cite{alpaca}        & Stanford      & 7  & 13B: 919 (22 / 28) & \checkmark & \checkmark & \checkmark & \checkmark \\
FastChat-t5-3b~\cite{fastt5}   & LMSYS         & 3  & 888 (25 / 28)      & \checkmark & \checkmark & \checkmark & \checkmark \\
ChatGLM-Pro~\cite{glm}      & Tsinghua      & /  & N/A                & \checkmark & \checkmark &   &   \\
Baichuan-2-13b~\cite{yang2023baichuan}  & Baichuan Inc. & 13 & N/A                & \checkmark & \checkmark &   &  \\
\bottomrule[1.0pt]
\end{tabular}
\end{table*}

In this section, we provide a quick introduction to the experimental setup,
 including the selection of tasks and LLMs, baseline setting, evaluation metrics, implementation details of the evaluation framework, and manual annotations.

\subsection{Tasks and LLMs Selection}
Given the limitations of multiple-choice questions, we chose two representative generation tasks that are more generalizable and more closely matched to real-world needs: text summarization and non-factoid question answering (QA).

As for the text summarization task, we adopted Extreme Summarization (XSum)~\cite{xsum} dataset to construct evaluation tasks. XSum is a real-world single-document news summary dataset collected from online articles by the British Broadcasting Corporation (BBC) and has been widely used in previous research~\cite{chowdhery2022palm, li2021prefix}. The entire XSum dataset contains over 220 thousand news documents.

As for the non-factoid QA task, we used the NF-CATS dataset~\cite{bolotova2022nfcats} to create evaluation tasks. NF-CATS is an emerging non-factoid QA dataset that contains 11,984 non-factoid questions as well as their categorizations. We removed the questions belonging to the types ``FACTOID'' and ``NOT-A-QUESTION'' to construct the sample pooling set.

To validate the effectiveness of each evaluation method, we need to collect the most reliable evaluation data, i.e., human preferences over the LLM's outputs for each test case, as our ground truth. Due to our limited budget, we randomly sampled 100 samples from the XSum and NF-CATS datasets and used them as our testbed. 
 In our experiments, we did not choose multiple-choice-format tasks like SuperCLUE~\cite{superclue}, because they impose restrictions on the outputs of LLMs to be a particular label or option, making the evaluation approaches based on these tasks not generalizable to others. Instead, we directly asked evaluatee LLMs to provide response to each task and used the reviewer LLMs to rate the evaluatee's response, both in pointwise and pairwise manners.

We selected eleven powerful LLMs to conduct experiments, including LLMs in both closed-source (e.g. GPT-4 and Claude-1) and open-source (e.g. Llama-2-70b-chat and RWKV-4-Raven-7B) settings. In our experiments, these LLMs play dual roles as both evaluatees and reviewer candidates. Table~\ref{tab:The basic information of the large language models used in our experiments} shows some basic information about these LLMs, as well as their ratings and rankings in the ELO leaderboard (i.e., a leaderboard of LLMs created based on human annotations) of LMSYS released in September 2023~\cite{zheng2023judging}. GPT-4 and Claude-1, recognized as two of the strongest existing LLMs, are ranked in the top two positions on the ELO leaderboard.

\subsection{Baselines}
We compare the performance of the PRE model with several baselines, including: 

\textbf{ROUGE~\cite{rouge}, BLEU~\cite{bleu}, and BERTScore~\cite{bertscore}}: They are all reference-based word similarity metrics.
These are a series of reference-based word similarity metrics. Both ROUGE and BLEU are classic unsupervised text similarity metrics. In our experiments, we use three variants of ROUGE (ROUGE-1, ROUGE-2 and ROUGE-L) as well as two variants of BLEU (BLEU-1 and BLEU-2). BERTScore, on the other hand, adopts a pretrained BERT-like model as the cornerstone and evaluates the similarity between the contextual representation of generative text and reference text. Here, we use deberta-xlarge-mnli~\cite{he2020deberta} and roberta-large~\cite{liu2019roberta} as the base models for BERTScore. 

\textbf{PandaLM~\cite{pandalm}}: A fine-tuned language model based on Llama-7b~\cite{llama} for the preference judgment tasks.

\textbf{GPTScore~\cite{gptscore}}: Evaluate the quality of generative text based on its generation probability feeding into particular LLMs right after the given prompts.
 We use GPT-3 (text-davinci-003)~\cite{brown2020language} and FLAN-T5 (FT5-XL)~\cite{flant5} as the base models.

\textbf{Single LLM}: Only use a single LLM as an evaluator to assess the quality of the generative text. Its prompt setting is the same as the PRE model. The LLMs selected here are listed in Table~\ref{tab:The basic information of the large language models used in our experiments}.

\subsection{Meta-evaluation Metrics}
In our experiments, we collected manual annotations as the gold standard for the quality of LLM-generated summaries to evaluate the evaluation performance of the PRE model and baselines. Our annotation data has been organized into two different formats: (1) pairwise preferences, denoted as $z(s_1, s_2, t)$, which indicate the user preference between summary $s_1$ and $s_2$ in terms of summarization quality for text $t$. The function $z$ assigns values of either $s_1$ or $s_2$ to represent the better summary; (2) pointwise labels, denoted as $y(s,t)$, which indicate the quality of summary $s$ summarizing text $t$. The details on the collection of manual annotation will be discussed in Section~\ref{subsec:Manual Annotations}.

For these two different formats of labels, we proposed various evaluation metrics to measure the performance of LLM evaluation models. Specifically, (1) for pairwise labels, we use \textbf{Agreement} (A) to measure the proportion of identical preference results between the model and human annotations. (2) for pointwise labels, we use \textbf{Kendall's tau} ($\tau$)~\cite{kendall1938new} and \textbf{Spearman correlation coefficient} ($S$)~\cite{lehman2013jmp} to measure the consistency between the model's outputs $\hat{y}(s, t)$ and labels $y(s,t)$. 
 We calculate $\tau$ and $S$ for each task, and report the mean of them as the overall performance.

\subsection{Framework Details}
\label{subsec:framework details}
\subsubsection{Qualification exam module}
To test the ability of reviewer candidate LLMs, we selected the outputs (i.e., summaries of test documents) of three evaluatee LLMs with varying quality: GPT-3.5-turbo, Fastchat-t5-3b, and Alpaca-7b, as ``questioners''. Reviewer candidates are asked to rate these summaries. We designed three rating methods: \textbf{5-level pointwise}, \textbf{100-level pointwise}, and \textbf{pairwise} (or called \textbf{preference}). In both the 5-level and 100-level pointwise rating methods, the candidate LLMs need to rate an integer number for each (text, summary) pair to indicate its summarization quality. The differences between 5-level and 100-level settings are not only in the rating scale and granularity (1-5 levels and 0-100 levels), but also in the guidance style: the 5-level method offers detailed definition of each level, while the 100-level method only provides a general description on the quality tendency. The pairwise rating method requires candidate LLMs to rate the preference for each (text, summary 1, summary 2) tuple, determining which summary better summarize the text. To reduce bias caused by word position and frequency, we constructed two prompt samples ($(t,s_1,s_2)$ and $(t,s_2,s_1)$) for each text-summary-summary tuple $(t,s_1,s_2)$ in our experiments. 

We uniformly designed prompts for these three rating methods, as specified in the Appendix~\ref{sec:The Design of Evaluation Prompt}. Additionally, we collected human preferences as the ground truth for the exam, and then used Agreement (e.g., in the pointwise cases like 5-level or 100-level ratings, convert the rates to pairwise preferences first) in the pairwise mode as the evaluation metric to rate the evaluation ability of candidate LLMs. Only when an LLM's Agreement exceeds the threshold of $\xi$, it will be retained as a reviewer for the peer review process. In our experiments, we set $\xi$ to be $60\%$.

\subsubsection{Peer review module}
For the text summarization task, we have devised a unified set of prompts to be fed into the whole eleven evaluatee LLMs. Specifically, we utilized the prompt template ``\textit{Task: Generate a short summary of the text in at most 64 words. Text: \{original text\} Summary:}''. Then, only the LLMs that pass the qualification exam are deployed to rate the outputs of evaluatee LLMs. We fed the prompts designed in the Appendix~\ref{sec:The Design of Evaluation Prompt} into reviewer LLMs and collected their scoring results. Overall, in the pointwise and pairwise modes, each reviewer LLM is required to generate $11 \times 100 = 1,100$ rates or $Perm(11,2) \times 100 = 11,000$ preferences, respectively.

\subsubsection{``Chair'' decision module}
In Sec~\ref{subsec:chair decision module}, Eq~\ref{eq:weighted vote strategy} already demonstrates the core idea of the weighted voting strategy. For pointwise and pairwise modes, we have different implementation details:

For the pointwise mode, each text-summary pair is treated as a sample $x$. We first need to normalize the original LLM output score $r_x^{(l)}$ using mean-variance normalization to eliminate its weighting effect. The weight of reviewer LLM $w_l$ is determined by its Agreement $p_l$ in the qualification exam. In the experiments, we set $w_l = \log\left(\dfrac{p_l}{1-p_l}\right)$, just as Eq~\ref{eq:aggregation model on pointwise mode} shows, where $W$ is the normalization term.

\begin{equation}
	R_x = \dfrac{1}{W} \sum_{l \in L} w_l \tilde{r}_x^{(l)} = \dfrac{1}{W} \sum_{l \in L} \log\left(\dfrac{p_l}{1-p_l}\right) \dfrac{r_x^{(l)} - \mu_l}{\sigma_l}
	\label{eq:aggregation model on pointwise mode}
\end{equation}

For the pairwise mode, let $x$ represent a text-summary-summary tuple $(t_x, s_{1,x}, s_{2,x})$. Each reviewer LLM $l$ votes for its preference output $r_x^{(l)}$ (either $s_{1,x}$ or $s_{2,x}$) with weight $w_l$. The preference result of the aggregated PRE model is determined by the summary with the higher votes. In our experiments, we also set $w_l = \log\left(\dfrac{p_l}{1-p_l}\right)$, just as shown in Eq~\ref{eq:aggregation model on pairwise mode}. The function $I(\cdot)$ denotes as the 0-1 Indicator function.

\begin{equation}
	R_x = \argmax_{s \in \{s_{1,x}, s_{2,x} \}} \sum_{l\in L} \log\left(\dfrac{p_l}{1-p_l}\right) I(r_{x}^{(l)} = s)
	\label{eq:aggregation model on pairwise mode}
\end{equation}

\begin{table*}[]
\caption{The overall performance of our proposed PRE models and baselines, evaluated by \textbf{Agreement} metric. The bold text indicates the best performing model. $\dagger$/$\dagger \dagger$ indicates $p$-value of paired sample t-test where the method outperforms GPT-4 is less than 0.05/0.01. The methods in the above part of table are compared with GPT-4 under pairwise setting. The underlined text denotes that the LLM passes the pairwise / 5-level / 100-level qualification exam, respectively.} 
\label{tab:The overall performance of our proposed PRE models and baselines}
\begin{subtable}{.9\linewidth}
\centering
\caption{PRE and single LLM models}	
\begin{tabular}{l|cccccc}
\toprule[1.0pt]
\multicolumn{1}{c|}{\multirow{2}{*}{Evaluation Models}} &
  \multicolumn{3}{c|}{XSum} &
  \multicolumn{3}{c}{NF-CATS} \\
\multicolumn{1}{c|}{} &
  \multicolumn{1}{l}{pairwise} &
  \multicolumn{1}{l}{5-level point} &
  \multicolumn{1}{l|}{100-level point} &
  \multicolumn{1}{l}{pairwise} &
  \multicolumn{1}{l}{5-level point} &
  \multicolumn{1}{l}{100-level point} \\ \hline
RWKV-4-Raven-7B &
  0.4972 &
  0.0000 &
  \multicolumn{1}{c|}{0.0000} &
  0.5021 &
  0.5083 &
  0.4958 \\
Alpaca-7b &
  0.5056 &
  0.3249 &
  \multicolumn{1}{c|}{0.3940} &
  0.5286 &
  0.5455 &
  0.5155 \\
Vicuna-7b &
  0.4948 &
  0.4721 &
  \multicolumn{1}{c|}{0.4732} &
  0.5296 &
  0.5557 &
  0.5574 \\
ChatGLM2-6B &
  0.5619 &
  {\ul 0.5839} &
  \multicolumn{1}{c|}{{\ul 0.6135}} &
  0.5414 &
  0.5735 &
  0.5958 \\
Baichuan-2-13b &
  {\ul 0.6057} &
  0.5471 &
  \multicolumn{1}{c|}{0.5653} &
  0.5515 &
  0.5521 &
  0.5500 \\
Llama-2-70b-chat &
  {\ul 0.5719} &
  {\ul 0.5848} &
  \multicolumn{1}{c|}{{\ul 0.6704}} &
  {\ul 0.5891} &
  0.5515 &
  {\ul 0.6798} \\
GPT-3.5-turbo &
  {\ul 0.6470} &
  {\ul 0.6676} &
  \multicolumn{1}{c|}{{\ul 0.6361}} &
  {\ul 0.6080} &
  0.5586 &
  0.5592 \\
Claude-1 &
  {\ul 0.6729} &
  {\ul 0.6484} &
  \multicolumn{1}{c|}{{\ul 0.6467}} &
  {\ul 0.6613} &
  {\ul 0.5774} &
  {\ul 0.5881} \\
FastChat-t5-3b &
  {\ul 0.6921} &
  {\ul 0.6291} &
  \multicolumn{1}{c|}{{\ul 0.6302}} &
  {\ul 0.6537} &
  0.5411 &
  0.5708 \\
ChatGLM-Pro &
  {\ul 0.6951} &
  {\ul 0.6701} &
  \multicolumn{1}{c|}{{\ul 0.7158}} &
  {\ul 0.7042} &
  {\ul 0.6485} &
  {\ul 0.6887} \\
GPT-4 &
  {\ul 0.7369} &
  {\ul 0.6958} &
  \multicolumn{1}{c|}{{\ul 0.7206}} &
  {\ul 0.7815} &
  {\ul 0.6330} &
  {\ul 0.6801} \\ \hline
PRE w/o GPT-4 (ours) &
  0.7328 &
  0.7242$\dagger$ &
  \multicolumn{1}{c|}{0.7334} &
  0.7402 &
  0.6604$\dagger \dagger$ &
  0.7074$\dagger$ \\
PRE (ours) &
  \textbf{0.7443} &
  \textbf{0.7331}$\dagger \dagger$ &
  \multicolumn{1}{c|}{\textbf{0.7390}$\dagger \dagger$} &
  \textbf{0.7842} &
  \textbf{0.6935$\dagger \dagger$} &
  \textbf{0.7113$\dagger \dagger$} \\
\bottomrule[1.0pt]
\end{tabular}
\end{subtable}

\begin{subtable}{.9\linewidth}
\centering	
\caption{Other baseline models}
\begin{tabular}{l|cccccc}
\toprule[1.0pt]
Evaluation Models &
  XSum &
  \multicolumn{1}{c|}{NF-CATS} &
  \multicolumn{2}{c|}{models} &
  XSum &
  NF\_CATS \\ \hline
BERTScore (roberta) &
  0.5728 &
  \multicolumn{1}{c|}{/} &
  \multicolumn{2}{c|}{BLEU-1} &
  0.5505 &
  / \\
BERTScore (deberta) &
  0.5901 &
  \multicolumn{1}{c|}{/} &
  \multicolumn{2}{c|}{BLEU-2} &
  0.5558 &
  / \\
PandasLM &
  0.6350 &
  \multicolumn{1}{c|}{0.7205} &
  \multicolumn{2}{c|}{ROUGE-1} &
  0.5884 &
  / \\
GPTScore (flan-t5-xl) &
  0.6023 &
  \multicolumn{1}{c|}{0.4762} &
  \multicolumn{2}{c|}{ROUGE-2} &
  0.5636 &
  / \\
GPTScore (text-davinci-003) &
  0.6910 &
  \multicolumn{1}{c|}{0.5940} &
  \multicolumn{2}{c|}{ROUGE-l} &
  0.5798 &
  / \\ 
\bottomrule[1.0pt]
\end{tabular}
\end{subtable}
\end{table*}
\begin{table*}[]
\caption{The overall performance of our proposed PRE models and baselines, evaluated by pointwise metrics, i.e., \textbf{Kendall's tau} ($\tau$) and \textbf{Spearman correlation coefficient} (S). The bold text indicates the best performing model. $\dagger$/$\dagger \dagger$ indicates $p$-value of paired sample t-test where the method outperforms GPT-4 is less than 0.05/0.01. The methods in the above part of table are compared with GPT-4 under 5-level point setting.}
\label{tab:overall performance appendix}
\centering

\begin{subtable}{.9\linewidth}
\centering
\caption{PRE and single LLM models}
\begin{tabular}{l|cccc|cccc}
\toprule[1.0pt]
\multicolumn{1}{c|}{\multirow{3}{*}{models}} &
  \multicolumn{4}{c|}{XSum} &
  \multicolumn{4}{c}{NF-CATS} \\
\multicolumn{1}{c|}{} &
  \multicolumn{2}{c}{5-level point} &
  \multicolumn{2}{c|}{100-level point} &
  \multicolumn{2}{c}{5-level point} &
  \multicolumn{2}{c}{100-level point} \\
\multicolumn{1}{c|}{} &
  \multicolumn{1}{c}{$\tau$} &
  \multicolumn{1}{c}{S} &
  \multicolumn{1}{c}{$\tau$} &
  \multicolumn{1}{c|}{S} &
  \multicolumn{1}{c}{$\tau$} &
  \multicolumn{1}{c}{S} &
  \multicolumn{1}{c}{$\tau$} &
  \multicolumn{1}{c}{S} \\ \hline
RWKV-4-Raven-7B &
  / &
  / &
  / &
  / &
  0.0277 &
  0.0482 &
  -0.0277 &
  -0.0334 \\
Alpaca-7b &
  0.0335 &
  0.0500 &
  0.0306 &
  0.0506 &
  0.0489 &
  0.0856 &
  0.0250 &
  0.0390 \\
Vicuna-7b &
  -0.0028 &
  -0.0135 &
  0.0175 &
  0.0330 &
  0.0854 &
  0.1738 &
  0.0925 &
  0.1512 \\
ChatGLM2-6B &
  0.1305 &
  0.2268 &
  0.1406 &
  0.2172 &
  0.0990 &
  0.1664 &
  0.1394 &
  0.2333 \\
Llama-2-70b-chat &
  0.1386 &
  0.3079 &
  0.2628 &
  0.4503 &
  0.0950 &
  0.1908 &
  0.2184 &
  0.3576 \\
Baichuan-2-13b &
  0.0941 &
  0.2255 &
  0.1185 &
  0.2271 &
  0.0865 &
  0.1735 &
  0.0833 &
  0.1381 \\
GPT-3.5-turbo &
  0.2495 &
  0.4199 &
  0.2029 &
  0.3165 &
  0.0757 &
  0.1577 &
  0.0929 &
  0.1520 \\
Claude-1 &
  0.2491 &
  0.4650 &
  0.2702 &
  0.4321 &
  0.1023 &
  0.1949 &
  0.0795 &
  0.1355 \\
FastChat-t5-3b &
  0.2090 &
  0.3935 &
  0.2195 &
  0.3295 &
  0.0638 &
  0.1439 &
  0.1107 &
  0.1716 \\
ChatGLM-Pro &
  0.2662 &
  0.4898 &
  0.3129 &
  0.4868 &
  0.2038 &
  0.3605 &
  0.2357 &
  \textbf{0.3893} \\
GPT-4 &
  0.3098 &
  0.4929 &
  0.3290 &
  0.4845 &
  0.1776 &
  0.3318 &
  0.2052 &
  0.3287 \\ \hline
PRE w/o GPT-4 (ours) &
  0.3271 &
  0.4835 &
  0.3052 &
  0.4543 &
  0.2043 &
  0.3451 &
  0.2329 &
  0.3601 \\
PRE (ours) &
  \textbf{0.3452$\dagger \dagger$} &
  \textbf{0.4998} &
  \textbf{0.3319} &
  \textbf{0.4947} &
  \textbf{0.2348} &
  \textbf{0.3843} &
  \textbf{0.2438} &
  0.3735 \\
\bottomrule[1.0pt]
\end{tabular}
\end{subtable}

\begin{subtable}{.9\linewidth}
\centering
\caption{Other baseline models}	
\begin{tabular}{l|cccc|cccc}
\toprule[1.0pt]
\multicolumn{1}{c|}{\multirow{2}{*}{models}} &
  \multicolumn{2}{c}{XSum} &
  \multicolumn{2}{c|}{NF-CATS} &
  \multicolumn{2}{c|}{\multirow{2}{*}{models}} &
  \multicolumn{2}{c}{XSum} \\
\multicolumn{1}{c|}{} &
  $\tau$ &
  S &
  $\tau$ &
  S &
  \multicolumn{2}{c|}{} &
  tau &
  S \\ \hline
BERTScore (roberta) &
  \multicolumn{1}{l}{0.1562} &
  \multicolumn{1}{l}{0.2379} &
  / &
  / &
  \multicolumn{2}{c|}{BLEU-1} &
  \multicolumn{1}{l}{0.1371} &
  \multicolumn{1}{l}{0.1969} \\
BERTScore (deberta) &
  \multicolumn{1}{l}{0.1829} &
  \multicolumn{1}{l}{0.2715} &
  / &
  / &
  \multicolumn{2}{c|}{BLEU-2} &
  \multicolumn{1}{l}{0.1252} &
  \multicolumn{1}{l}{0.1864} \\
PandaLM &
  / &
  / &
  / &
  / &
  \multicolumn{2}{c|}{ROUGE-1} &
  \multicolumn{1}{l}{0.1662} &
  \multicolumn{1}{l}{0.2448} \\
GPTScore (flan-t5-xl) &
  \multicolumn{1}{l}{0.1486} &
  \multicolumn{1}{l}{0.2286} &
  \multicolumn{1}{l}{-0.0048} &
  \multicolumn{1}{l|}{0.0033} &
  \multicolumn{2}{c|}{ROUGE-2} &
  \multicolumn{1}{l}{0.1214} &
  \multicolumn{1}{l}{0.1789} \\
GPTScore (text-davinci-003) &
  \multicolumn{1}{l}{0.2848} &
  \multicolumn{1}{l}{0.4203} &
  \multicolumn{1}{l}{0.1238} &
  \multicolumn{1}{l|}{0.1966} &
  \multicolumn{2}{c|}{ROUGE-l} &
  \multicolumn{1}{l}{0.1524} &
  \multicolumn{1}{l}{0.2329} \\
\bottomrule[1.0pt]
\end{tabular}
\end{subtable}
\end{table*}

\subsection{Manual Annotations}
\label{subsec:Manual Annotations}
We conducted manual annotations serving for two purposes: (1) as ground truth for the LLM qualification exam (only use a small subset of annotations); (2) as a gold standard for evaluating the performance of different evaluation methods. Due to cost considerations, we conducted annotations on 7 out of the 11 LLMs that diversify in quality, developers, and model structure, as shown in Table~\ref{tab:The basic information of the large language models used in our experiments}.

To meet the requirements of both pointwise and pairwise evaluation metrics, we conducted pointwise annotation as well as auxiliary preference annotation \footnote{Due to the high cost, we conducted preference annotation only on the pairs with a tied pointwise label.}.

Here let us use the XSum dataset as an instance to introduce the annotation details. The design of NF-CATS dataset is quite similar to it.
In the first step, we recruited annotators to conduct pointwise annotation. In each annotation task, we provided the annotators with a (text, summary) pair (or (question, answer) pair in non-factoid QA tasks), and they are required to give a rating on a 5-level scale. We adopted the Likert scale~\cite{jebb2021review} as the 5-level annotation rule. 
For the statement ``The summary text adequately and briefly summarizes the core meaning of the original text'' levels $1 \sim 5$ respectively represent annotator strongly-disagrees/disagrees/neutralizes/agrees/strongly-agrees with the above statement. 
Furthermore, for all the summary pairs (in the same task) with tied ratings in the 5-level pointwise annotation, we recruited annotators to conduct preference annotations. Similar to the Likert scale, we designed a 7-level annotation rule ranging from -3 to 3. To improve the annotation experience of assessors, they are allowed to give any real number within the range $[-3, 3]$, where levels $-3 \sim 3$ respectively represent ``compared to summary text 2, summary text 1 summarizes the core meaning of the original text strongly-better/better/slightly-better/tied/slightly-worse/worse/strongly-worse''. The difference compared with general 7-level setting is that the assessment results are allowed to be any real number within the range $[-3, 3]$ to improve the annotation experience of annotators.

We recruited annotators through Amazon's MTurk Crowdsourcing platform~\footnote{\url{https://www.mturk.com/}}, assigning 5 different annotators for each Human Intelligence Task (HIT). Overall, we collected 1,400 pointwise HITs and 1,704 preference HITs, collecting a total of 15,520 annotations at a cost of approximately 1,600 dollars. After removing the maximum and minimum labels, the annotations achieve fair annotation agreement: the mean intra-task Krippendorff's $\alpha$~\cite{krippendorff2011computing}  for pointwise and preference annotations are $0.4581$ and $0.2983$, respectively.

\section{Results and Analysis}
\label{sec:Results and Analysis}

In this section, we present the experimental results and attempt to answer the following three research questions (RQs): 

 \begin{enumerate}
 	\item How does the performance of our proposed PRE model compare to other baseline methods?
 	\item Does the inherit bias really exist when evaluating with a single LLM?
 	\item How robust is the PRE model in evaluating LLMs?
 \end{enumerate}

\vspace{-3mm}
\subsection{Overall Results (RQ1)}

Table~\ref{tab:The overall performance of our proposed PRE models and baselines} presents the overall experimental results, evaluated by Agreement metric, where \textit{PRE w/o GPT-4} represents the PRE variant model that excludes GPT-4 (currently recognized as the strongest LLM) out of the reviewer list. The results show that our proposed PRE model outperforms all the baselines including GPT-4. Even the variant without the GPT-4 model (\textit{PRE w/o GPT-4}) achieves comparable evaluation results with GPT-4. This indicates that the peer review mechanism could effectively evaluate.

Experimental results show that GPT-4 performs the best among all LLMs in terms of evaluating the outputs of evaluatee LLMs. GPT-3.5-turbo, Claude-1 and ChatGLM-Pro also perform well in the evaluation task, as we expect. Surprisingly, FastChat-t5-3b, a model with only 3 billion parameters, achieves a comparable evaluation level to larger-scale LLMs such as Claude-1 and ChatGLM-Pro when evaluating text summarization tasks. We speculate that this is caused by the detailed design of its instruct tuning strategy during training, which makes it effective in dealing with such specific rating tasks. By contrast, it performs relatively average in evaluating non-factoid QA tasks.

When comparing three different prompt settings, we find that the pairwise setting is slightly better than the pointwise ones, while the performance difference between the 5-level and 100-level pointwise settings is not significant. Therefore, we recommend using the pairwise setting when resources permit. 

Table~\ref{tab:The overall performance of our proposed PRE models and baselines} also shows the performance of reference-based word similarity metrics such as ROUGE, BLUE, and BERTScore. We find that these metrics have positive correlations with human annotations, but the overall evaluation performance is worse compared to LLM-based methods like GPT-4 and PRE. PandaLM and GPTScore (text-davinci-003) show competitive performance in NF-CATS and XSum tasks respectively, but not in the other one. This phenomenon shows their performance is not robust across different tasks.


 One point that needs clarification is that Table~\ref{tab:The overall performance of our proposed PRE models and baselines} shows that RWKV-4-Raven-7B and Alpaca-7b have an Agreement of preference prediction much less than 50\% under the 5-level and 100-level pointwise settings in XSum dataset. This is mainly because these two LLMs have difficulties in understanding the problems in many cases, leading to the failure on extracting useful rating information from their outputs.

 \begin{figure*}[thbp]
   \captionsetup[subfigure]{justification=centering}
   \centering
	\subfloat[Pairwise Setting]{
     \includegraphics[width=0.3\textwidth]{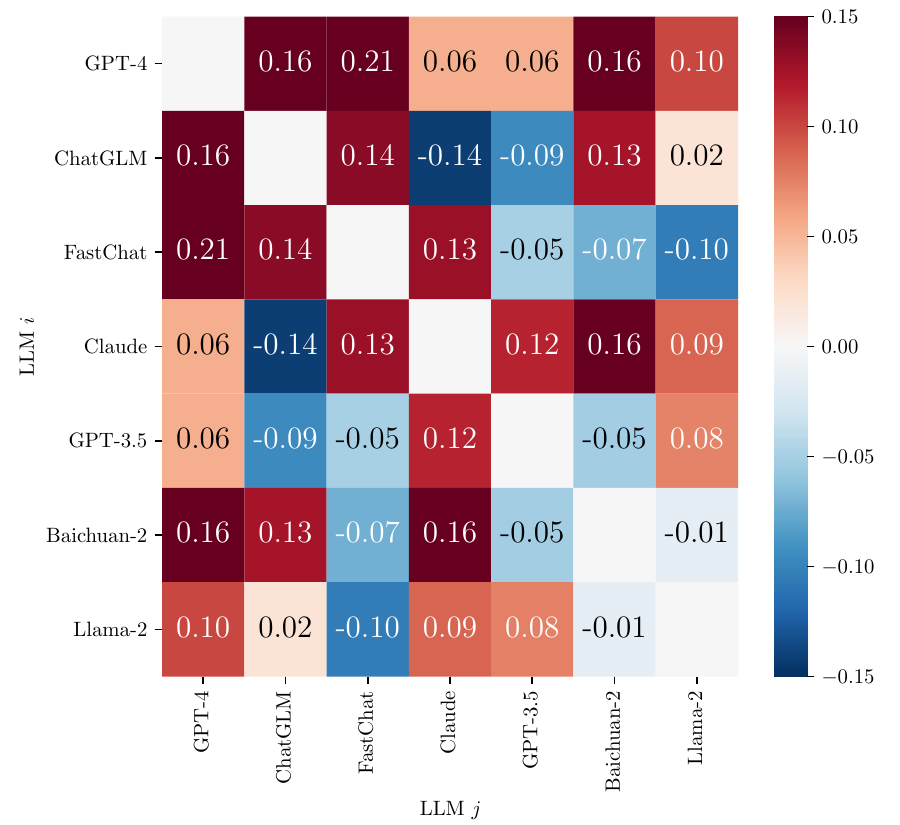}
     \label{fig:Pairwise Setting, biasness}
   } 
   \subfloat[5-level Pointwise Setting]{
     \includegraphics[width=0.3\textwidth]{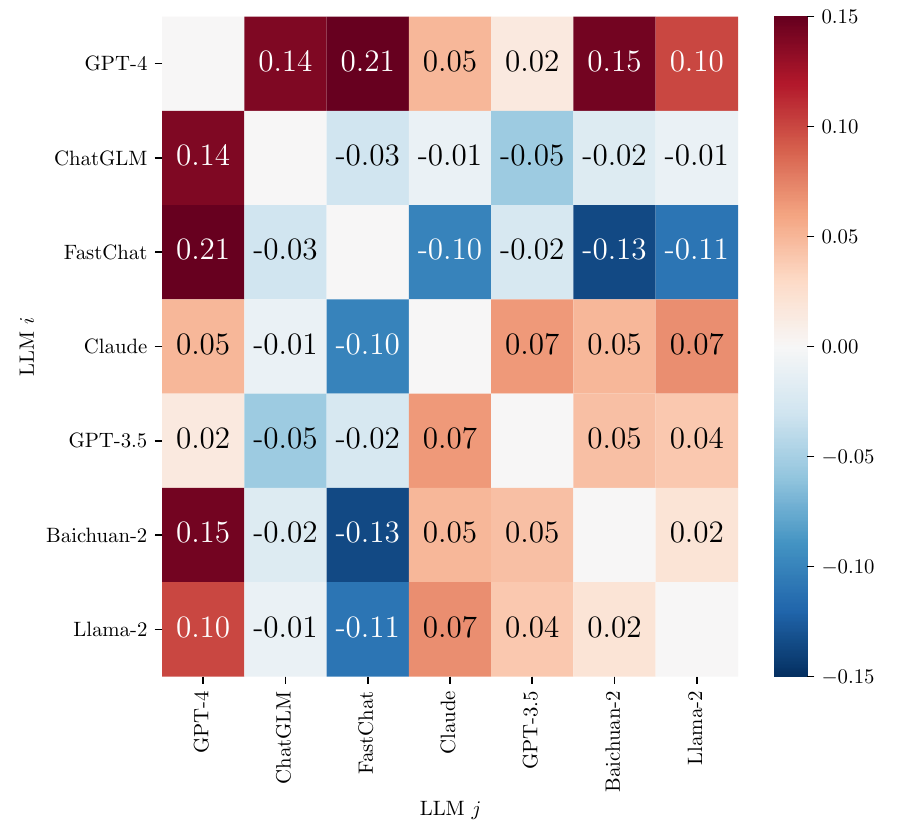}
     \label{fig:5-level Pointwise Setting, biasness}
   }
   \subfloat[100-level Pointwise Setting]{
     \includegraphics[width=0.3\textwidth]{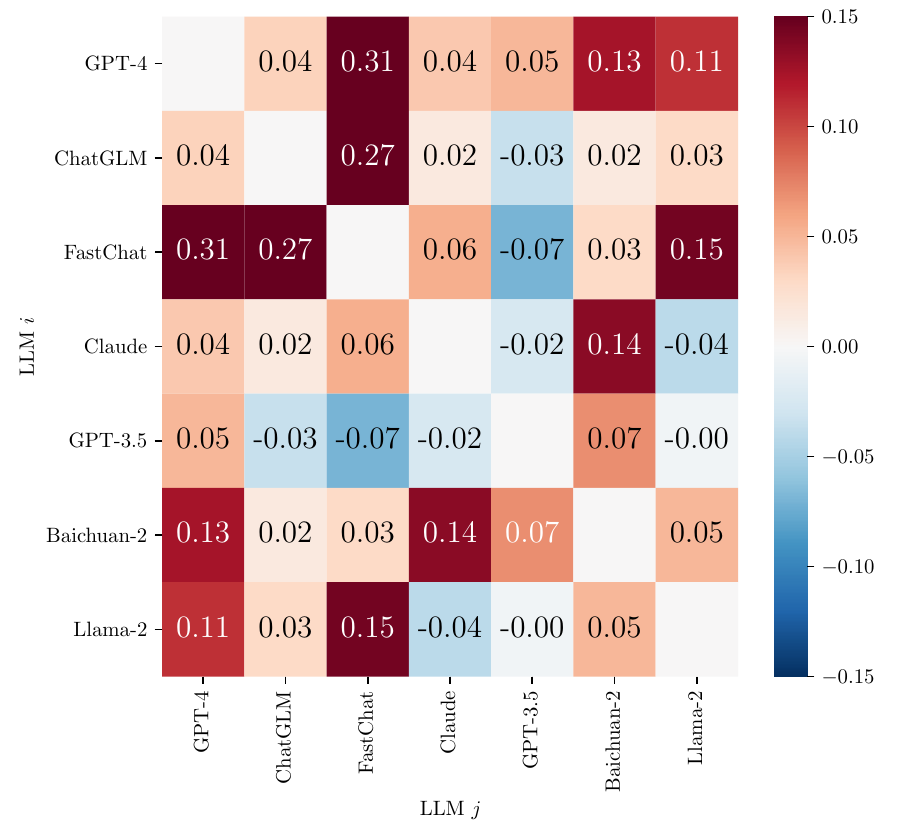}
     \label{fig:100-level Pointwise Setting, biasness}
   }
   \caption{The severity of bias among seven powerful LLMs (measured with metric Preference Gap). Larger value (greater than 0) indicates a higher potential for bias between those two LLMs.}
   \label{fig:The severity of biasness among sever powerful LLMs}
\end{figure*}

\begin{figure}[thbp]
   \captionsetup[subfigure]{justification=centering}
   \centering
 \includegraphics[width=0.45\textwidth]{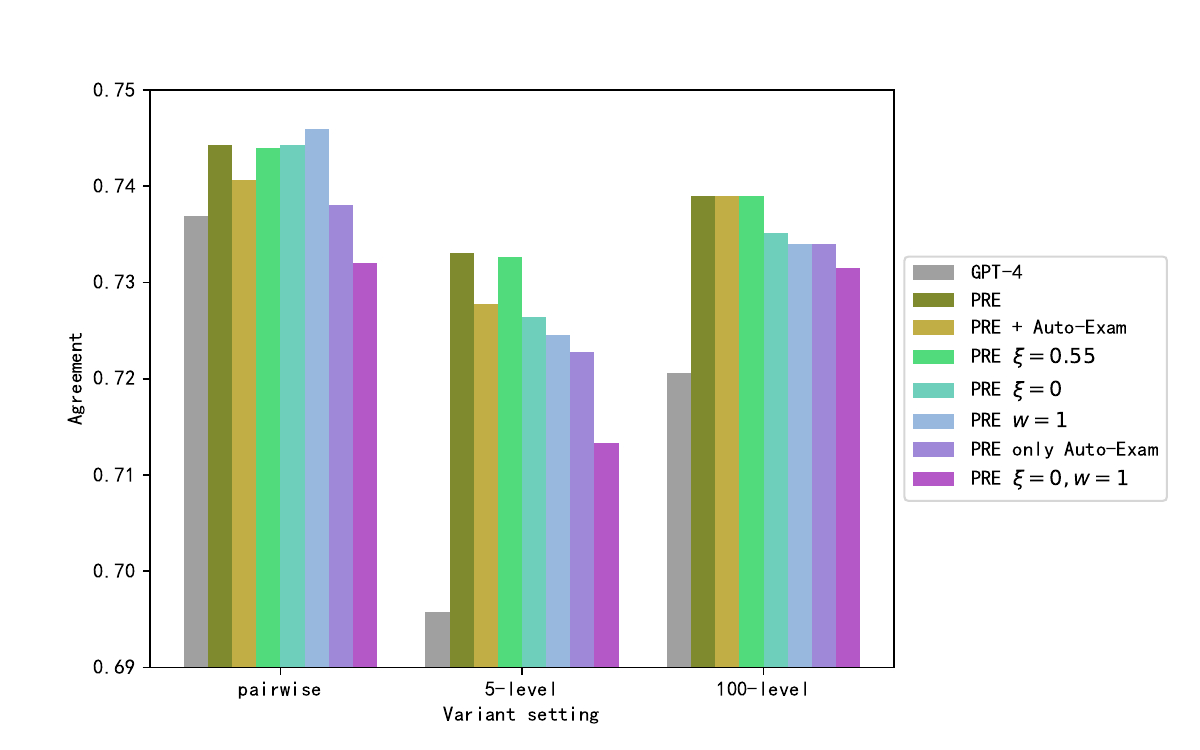}
   \caption{The performance of several PRE variants under different settings on XSum}
   \label{fig:The performance of several PRE variants under different prompt settings}
 \end{figure}

\vspace{-3mm}
\subsection{Bias Analysis (RQ2)}
\label{subsec:biasness analysis}
In this section, we dive into the results provided by different evaluation methods and investigate whether we could observe any type of evaluation bias in each of them. To measure potential bias in the evaluation using a single LLM, we propose the preference gap (PG) as an evaluation metric. Specifically, for LLMs $i$ and $j$, we define the preference gap between LLM $i$ and $j$ ($\text{PG}(i, j)$) as the proportion of $i$'s outputs that are better than $j$'s outputs from $i$'s perspective, subtracted by the same proportion from $j$'s perspective, as shown in Eq~\ref{eq:definition of the preference gap measure}. Naturally, the larger $\text{PG}(i, j)$ is, the more likely the bias exists between LLMs $i$ and $j$. Ideally, for models without bias, the distribution of PG in the set $\{PG(i, j) | i \in L, j \in L, i \neq j \}$ is a random noise with a mean of 0.

\begin{equation}
	\text{PG}(i, j) = P_i(i \succ j) - P_j(i \succ j)
	\label{eq:definition of the preference gap measure}
\end{equation}

We conducted experiments under XSum tasks. Figure~\ref{fig:The severity of biasness among sever powerful LLMs} shows the heatmap distribution of the PG metric among seven powerful LLMs 
under different settings in XSum tasks. In the pairwise, 5-level pointwise and 100-level pointwise settings, the proportions of PG values greater than 0 (i.e., $i$ has stronger preferences than $j$ on the output of $i$) are 66.67\%, 57.14\% and 76.19\% respectively, which are all significantly higher than the 50\% of the unbiased scenario. 
 Furthermore, we conducted paired samples t-tests, resulting in $p$-values of 0.038, 0.263, and 0.006 for these three settings, respectively. 
The results indicate significant bias in evaluation using individual LLMs under the pairwise and 100-level pointwise settings. 
 Looking back to Figure~\ref{fig:The severity of biasness among sever powerful LLMs}, we also observe some more detailed conclusions: GPT-4 is likely to exhibit the most severe bias, as its PG values with all the other LLMs are all greater than 0 under all three settings; Baichuan-2-13b and Claude-1 also show relatively strong bias; the other four LLMs show weaker bias.

\subsection{Robust Analysis (RQ3)}
In this section, we aim to explore the robustness of the PRE model, that is, whether it still performs well when hyperparameters and qualification methods vary.

%

Here, we mainly attempted to adjust two hyperparameters: the pass threshold ($\xi$) for the qualification exam and the weight ($w_l$) used during rating aggregation. 
We adjusted $\xi$ to 55\% and 0, where $\xi=0$ indicates all candidate reviewers are allowed to participate in the peer review process. We also adjusted $w_l$ to be $1$, which means all reviewers have equal rating weight. 

We also tried an unsupervised qualification exam method called \textit{Auto-Exam}, in which we evaluated the consistency of the LLM outputs before and after changing the order of content in the prompt. 
 The output consistency is computed as the consistent proportion of the preference relations between two summaries of the same original text (under pointwise settings, ratings need to be converted into preference relations first for comparison). 
When the consistent proportion of such LLM exceeds a threshold $\eta$, this LLM is regarded as a reliable one to join in the reviewer set. 
 In our experiments, we adjusted the order of summary 1 and summary 2 under the pairwise setting, while adjusted the order of the original text and summary under the pointwise setting. Regarding the threshold, 
In our experiments, we set $\eta = 55\%$.

We conducted experiments in XSum tasks, Figure~\ref{fig:The performance of several PRE variants under different prompt settings} shows the performance of the PRE model under different hyperparameter and qualification settings, with GPT-4 used as the baseline. \textit{PRE + Auto-Exam} denotes the variant of PRE method with both the original exam and Auto-Exam, while \textit{PRE only Auto-Exam} denotes the variant with only Auto-Exam as the qualification exam and $w_l = 1$. Results show that the performance of the PRE model is not sensitive to the changes in its hyperparameters. Only when we remove all the effects of the qualification exam (i.e., $\xi=0, w_l=1$), does the performance of PRE noticeably decrease. This finding corroborates the necessity of LLM qualification filtering.

Figure~\ref{fig:The performance of several PRE variants under different prompt settings} also shows the effect of \textit{Auto-Exam} method. We find that PRE with only \textit{Auto-Exam} outperforms the non-exam one ($\xi=0,w=1$), but its performance is lower than the qualification exam with a subset of manual annotation as ground truth. This finding indicates the potential of \textit{Auto-Exam}, which deserves further exploration.

\vspace{-4mm}
\section{Conclusion}
\label{sec:conclusion}

In this paper, we propose a novel framework, Peer Review Evaluator (PRE), for automatically evaluating the performance of large language models (LLMs). Inspired by the peer-review mechanism in the academic community, we introduce a mutual evaluation mechanism among LLMs in our framework. By setting reasonable qualification exams and model aggregation criteria, our PRE model outperforms all baseline methods including GPT-4. In the experiments, we also validate the existence of bias when using a single model like GPT-4 as an evaluation tool. PRE could reduce this bias to some extent. We believe that our proposed PRE, an automatic LLM evaluation method, can be adaptable to various evaluation tasks and scenarios.

%
%
%
%
%
%
%
%

\appendix

\section{The Design of Evaluation Prompt}
\label{sec:The Design of Evaluation Prompt}
The evaluation prompts are adopted for both \textit{qualification exam} and \textit{peer review} modules (detailedly introduced in Sec~\ref{subsec:framework architecture}). These prompts are fed to the reviewer (or reviewer candidate) LLMs, allowing them to generate ratings or preferences. In our experiments, we have proposed three different prompt settings (pairwise, 5-level pointwise and 100-level pointwise), and then seperately designed the prompt template for each setting, as the following shows. Here we show the design in XSum dataset under each setting, the design of NF-CATS dataset is almost the same.
\subsection{Pairwise setting}

\begin{shaded}
\#\#\#Task: Evaluate two summaries of a given passage and determine which one better summarizes the main points of the passage considering accuracy and conciseness. You only need to output `one` or `two` directly to indicate which summary summarizes the passage better.\\~\\
\#\#\#Passage: \{ \textit{passage} \} \\~\\
\#\#\#Summary one: \{ \textit{summary 1} \} \\~\\
\#\#\#Summary two: \{ \textit{summary 2} \} \\~\\
\#\#\#Output:
\end{shaded}

\subsection{5-Level pointwise setting}
\begin{shaded}
\#\#\#Task: Evaluate the summary of a given passage and determine how it summarizes the main points of the passage considering accuracy and conciseness. Directly output a number between 1 and 5 to indicate the quality score of this summary: \\
- 1 means the summary is not relevant to the passage, \\
- 2 means the summary is neither accurate nor concise but it is relevant to the passage, \\
- 3 means the summary is only a fair summary of the passage considering accuracy and conciseness, \\
- 4 means the summary is a good summary of the passage but still has room for improvement in accuracy and conciseness, \\
- 5 means the summary is a perfect summary of the passage considering accuracy and conciseness. \\~\\
\#\#\#Passage: \{ \textit{passage} \} \\~\\
\#\#\#Summary: \{ \textit{summary} \} \\~\\
\#\#\#Score of the summary: 	
\end{shaded}

\subsection{100-Level pointwise setting}
\begin{shaded}
\#\#\#Task: Evaluate the summary of a given passage and determine how it summarizes the main points of the passage considering accuracy and conciseness. Directly output a number between 0 and 100 to indicate the score of this summary. The higher the score, the more accurate and concise the summary is. \\~\\
\#\#\#Passage: \{ \textit{passage} \} \\~\\
\#\#\#Summary: \{ \textit{summary} \} \\~\\
\#\#\#Score of the summary:	
\end{shaded}

\newpage
\bibliographystyle{ACM-Reference-Format}
\bibliography{main}

\end{document}